\documentclass[aps,prl,twocolumn,superscriptaddress,showpacs]{revtex4}
\usepackage{amsmath}
\usepackage{amssymb}
\usepackage{graphicx}

\begin{document}
\title{$\pi$-phases in balanced fermionic superfluids in spin-dependent optical lattices}
\author{I. Zapata}
\affiliation{Departamento de F\'{\i}sica de Materiales, Universidad Complutense de Madrid,
E-28040 Madrid, Spain}
\author{B. Wunsch}
\author{N.~T. Zinner}
\author{E. Demler}
\affiliation{Department of Physics, Harvard University, 17 Oxford Street, Cambridge, MA 02138, USA}

\pacs{67.85.-d,03.75.Ss,71.10.Pm,74.45.+c}
\date{\today}

\begin{abstract}
We study a balanced two-component system of ultracold fermions in one dimension
with attractive interactions and subject to a spin-dependent optical
lattice potential of opposite sign for the two components. 
We find states with different types of modulated pairing order parameters which are conceptually similar to $\pi$-phases discussed for superconductor-ferromagnet heterostructures.  Increasing the lattice depth induces sharp transitions between states of different parity. While the origin of the order paramter oscillations is similar to the FFLO phase for paired states with spin imbalance, the current system is intrinsically stable to phase separation. We discuss experimental requirements for creating and probing these novel phases. 
\end{abstract}
\volumeyear{2009}
\volumenumber{number}
\issuenumber{number}
\eid{identifier}
\startpage{1}
\endpage{ }
\maketitle

One of the most intriguing examples of the interplay of superconductivity and magnetism is the Fulde-Ferrel-Larkin-Ovchinnikov (FFLO) phase, where Zeeman splitting of the Fermi surfaces is expected to lead to spatial oscillations of the pairing amplitude. It is difficult to obtain such phases in superconductors, since the orbital effect of the magnetic field is typically much larger than the spin Zeeman splitting.  Several proposals have been made, however they remain controversial \cite{buzdin2005}. For example, FFLO phase has been discussed in the context of heavy fermion CeCoIn$_5$ superconductors \cite{Ce1,Ce2}, but alternative interpretation in terms of competing magnetic order has also been given \cite{Ce3}. So far the only unambiguous demonstration of FFLO-like physics has been achieved in heterostructures of ferromagnetic and superconducting (F/SC) layers \cite{Barsic2007}, where proximity coupling through ferromagnetic layers results in superconducting $\pi$-junctions (see \cite{buzdin2005} for a review). We note that $\pi$-phases arising from a different mechanism than FFLO have also been discussed for high-$T_c$ cuprates \cite{buzdin2005,berg2009}.

\begin{figure*}[htb!]
\includegraphics[width=2\columnwidth]{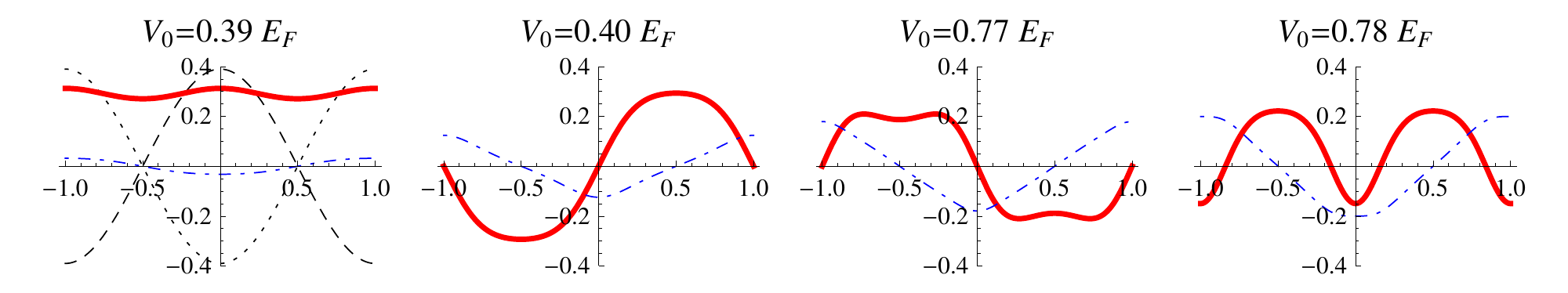}
\caption{The polarization $(n_{\uparrow}(x)-n_{\downarrow}(x))/k_F$ in dot-dashed blue and pairing $\Delta (x)/E_F$ in thick red as functions of $2x/\lambda$, for the case of 0-phase (left), $\pi$-phase (two figures in the middle), and $2\pi$-phase (right) at zero temperature. The dashed and dotted black lines in the left plot show the spin-dependent lattice potential. Here $k_F\lambda=30, T=0, g_{1D}k_F/E_F=-2.04$, and $V_{\uparrow}(x)=-V_{\downarrow}(x)=V_0 \cos (2 \pi x / \lambda)$.}
\label{figGapDensities}
\end{figure*}

Recently, in cold atoms, there has been a large body of work, both experimental and theoretical, aimed at achieving FFLO states. The biggest difficulty is that FFLO phases are fragile and extremely susceptible to phase separation and the experimental situation remains unclear \cite{bedaque2003,zwierlein2006,shin2006,schunck2007,partridge2006,hulet2009}.  In this paper we propose a novel system of ultracold fermions in an optical lattice \cite{jordens2008,schneider2008} which can be used to observe FFLO type states with oscillating pairing amplitude. The system which we discuss is somewhat similar to F/SC heterostructures and should be stable against phase separation. Our proposal relies on the ability to create spin dependent optical lattices and we find that beyond a certain critical strength of such optical potential, the superconducting pairing amplitude becomes a sign changing function (we will refer to such states as as $\pi$-phases, see Fig. \ref{figGapDensities}).

\begin{figure}[htb!]
\includegraphics[width=\columnwidth]{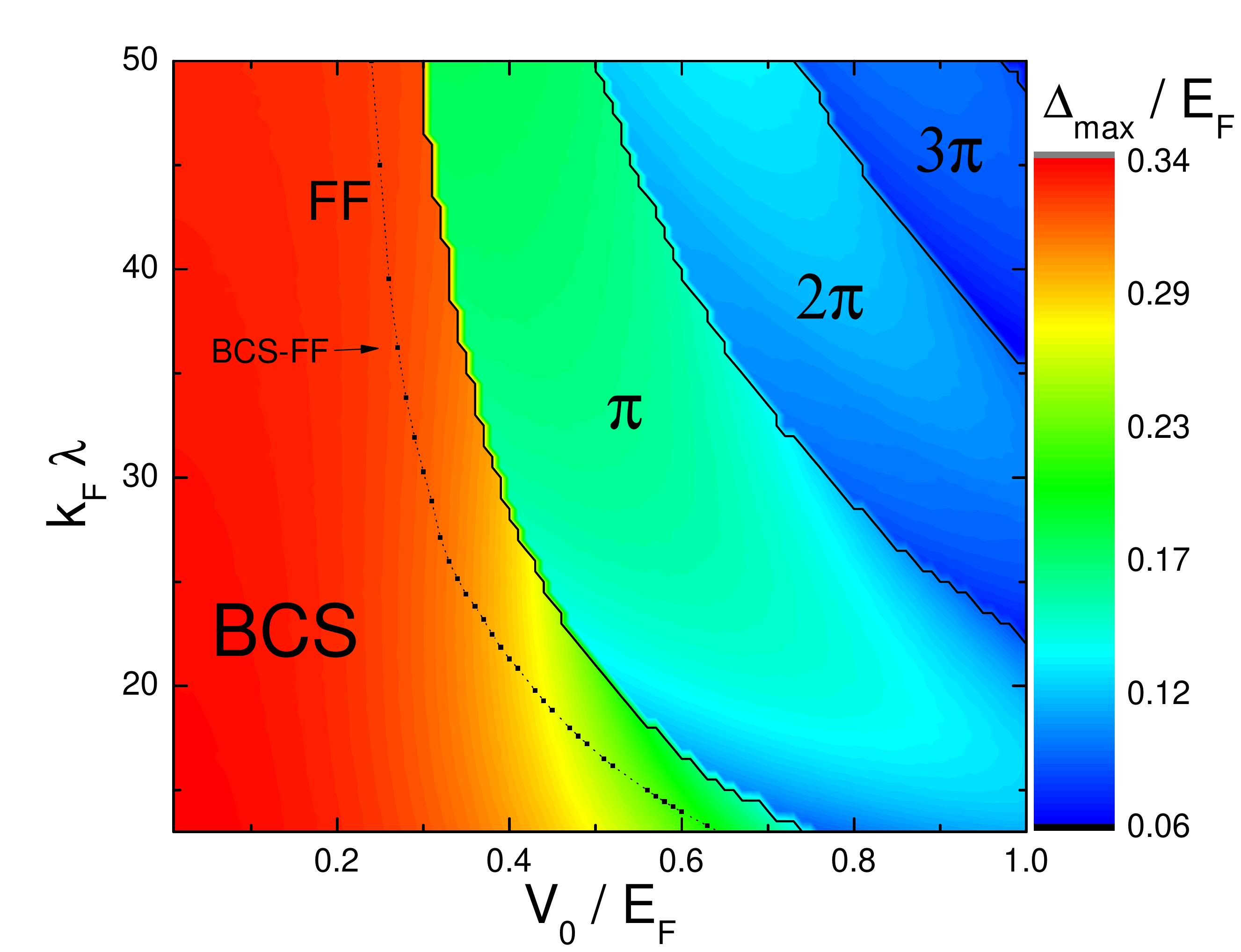}
\caption{Phase diagram showing the emergence of $\pi$-phases for a spin-dependent lattice potential of wavelength $\lambda$ and strength $V_0$ for interaction strength $g_{1D}k_F/E_F=-2.04$
and zero temperature. A gradient of colors gives $\Delta_{\text{max}}:=\text{max}|\Delta_{\tilde m}|$, the largest
Fourier component amplitude of the gap.
The black lines indicate transitions from gap profiles with zero (BCS), two ($\pi$), four ($2\pi$) and six ($3\pi$) zero-crossings per unit cell. The dashed FF line and the BCS-FF arrow are from a Fulde-Ferrell calculation in the homogeneuos system and is explained in the text.
}
\label{figPhaseDiagram}
\end{figure}

The gap profile in the ground-state depends
on the wavelength of the lattice, $\lambda$, the strength of the potential, $V_0$, and the
interaction strength. In
Fig.~\ref{figPhaseDiagram} we present the $(V_0,\lambda)$ phase diagram showing
the transitions from constant gap to the $\pi$-phases with several
zero-crossings per unit cell in the pairing amplitude.
A color gradient gives the largest Fourier component amplitude of the gap (see below)
and the black lines indicate the transitions.
We clearly
see $\pi$-phases occurring in a broad range of $\lambda$ restricted from below only by the
coherence length as we will discuss. The emergence of oscillations in the gap gives
clear signatures in the Fourier transform. We will demonstrate how the rapid-ramp
techniques can be used to observe these states in  time-of-flight measurements. We also
suggest ways to make spin dependent large wavelength lattice
potentials in the high-field regime as is needed to access $\pi$-phases.

The quasi-1D system we study is described by the effective Hamiltonian
\cite{drummond2007}
\begin{eqnarray}\label{eqnHamiltonian}
\nonumber
H &-& \mu_{\downarrow} N_{\downarrow} - \mu_{\uparrow} N_{\uparrow} = \\
\nonumber
& &\sum_{\sigma=\uparrow\downarrow} \int dx \Psi_{\sigma}^{\dag}(x)[-\frac{\hbar^2}{2m}\frac{\partial^2}{\partial x^2}+V_{\sigma}(x)-\mu_{\sigma}]\Psi_{\sigma}(x) \\
 &+& g_{1D}\int dx \Psi_{\uparrow}^{\dag}(x)\Psi_{\downarrow}^{\dag}(x)\Psi_{\downarrow}(x)\Psi_{\uparrow}(x),
\end{eqnarray}
where
$g_{1D}$ is the effective 1D coupling constant. We use $g_{1D}k_F/E_F=-2.04$ as in \cite{drummond2007} corresponding to an
interaction to kinetic energy ratio of $-m g_{1D}/\hbar^2=1.6$. This is neither weak coupling nor the unitary limit.

We consider a balanced system but introduce chemical potentials $\mu_{\sigma}$ since the optical lattice potential is spin-dependent. For the main part of this work we use $V_{\uparrow}(x)=V_{0} \cos (2\pi x/\lambda)$ and $V_{\downarrow}(x)=-V_{\uparrow}(x)$ in which case $\mu_{\uparrow}=\mu_{\downarrow}$.
In the non-interacting system, the spin-dependent lattice
spatially displaces the degenerate solution of the two components as $V_0$ is increased. In a simple-minded picture, pairing of these states will generate spatial variation in the order parameter which is the origin of $\pi$-phases.

We solve the Hamiltonian of Eq.~\eqref{eqnHamiltonian} in mean-field theory by using the inhomogeneous Bogoliubov-deGennes (BdG) ansatz for the field operator $\Psi_{\sigma}(x,t)=\sum_{k}[u_{k \sigma}(x)e^{-i \omega_{k \sigma} t} c_{k \sigma}+\sigma
\bar{v}_{k \bar{\sigma}}(x)e^{i \omega_{k \bar{\sigma}} t} c_{k \bar{\sigma}}^{\dag}]$, where the $c$ and $c^\dag$ denote the
quasiparticles and the sum runs over $\omega_{k \sigma}>0$ with $k$ the quasiparticle index composed of a quasimomentum and the band index.
The mean-field equations are
\begin{eqnarray}\label{eqnBdG}
\nonumber
  &&\left[
      \begin{array}{cc}
        H_{\sigma} & \Delta(x) \\
        \bar{\Delta}(x) & -H_{-\sigma} \\
      \end{array}
    \right]
    \left(
      \begin{array}{c}
        u_{k \sigma}(x) \\
        v_{k \sigma}(x) \\
      \end{array}
    \right) = \omega_{k \sigma}
    \left(
      \begin{array}{c}
        u_{k \sigma}(x) \\
        v_{k \sigma}(x) \\
      \end{array}
    \right), \\
  &&H_{\sigma} = -\frac{\hbar^2}{2m}\frac{\partial^2}{\partial x^2}+V_{\sigma}(x)-\mu_{\sigma}+g_{1D}n_{-\sigma}(x),
\end{eqnarray}
where $n_{\sigma}(x)=\langle\Psi_{\sigma}^{\dag}(x)\Psi_{\sigma}(x)\rangle$ and $\Delta(x)=-g_{1D}\langle\Psi_{\downarrow}(x)\Psi_{\uparrow}(x)\rangle$. These
equations can be solved self-consistently for densities and gap through
\begin{eqnarray}\label{eqnBdGSC}
\nonumber
n_{\uparrow}(x)&=&\sum_{\omega_{k \uparrow}} f(\omega_{k \uparrow}) |u_{k \uparrow}(x)|^2 =\sum_{\tilde m=-\infty}^{\infty} n_{\uparrow \tilde m}e^{i 2\pi\tilde m x/\lambda}\\
\nonumber
n_{\downarrow}(x)&=&\sum_{\omega_{k \uparrow}} f(-\omega_{k \uparrow}) |v_{k \uparrow}(x)|^2 =\sum_{\tilde m=-\infty}^{\infty} n_{\downarrow \tilde m}e^{i 2\pi \tilde m x/\lambda}\\
\nonumber
\Delta(x)&=&g_{1D} \sum_{\omega_{k \uparrow}} f(\omega_{k \uparrow}) u_{k \uparrow}(x)\bar{v}_{k \uparrow}(x)\\
&=&\sum_{\tilde m=-\infty}^{\infty} \Delta_{\tilde m}e^{i 2\pi \tilde m x/\lambda},
\end{eqnarray}
where $f(\omega_{k \sigma})=1/(1+\exp(\hbar \omega_{k \sigma}/k_B T))$ is the Fermi-Dirac distribution. In Eq.~\eqref{eqnBdGSC} we use
the periodicity of the optical lattice to do a Fourier decomposition.
Notice that these equations only contain $u,v$ for $\sigma=\uparrow$ and the sums over $\omega_{k\uparrow}$ are unrestricted \cite{drummond2007}.
We have explicitly checked the convergence of our numerical solutions by extending the cut-off on the basis size.

The mean-field BdG ansatz does not take
into account soft collective modes of the order parameter which, in principle, lead to the power law decay of the superconducting correlations. However, the BdG approach describes the ground state energy in our parameter regime well, which is determined by correlations on the scale of the BCS correlation length, $\xi=\hbar v_F/\Delta$ \cite{hu2007,drummond2007}. Thus we expect it to also correctly capture the
competition between $0$- and $\pi$-phases.

\begin{figure}
\includegraphics[width=\columnwidth]{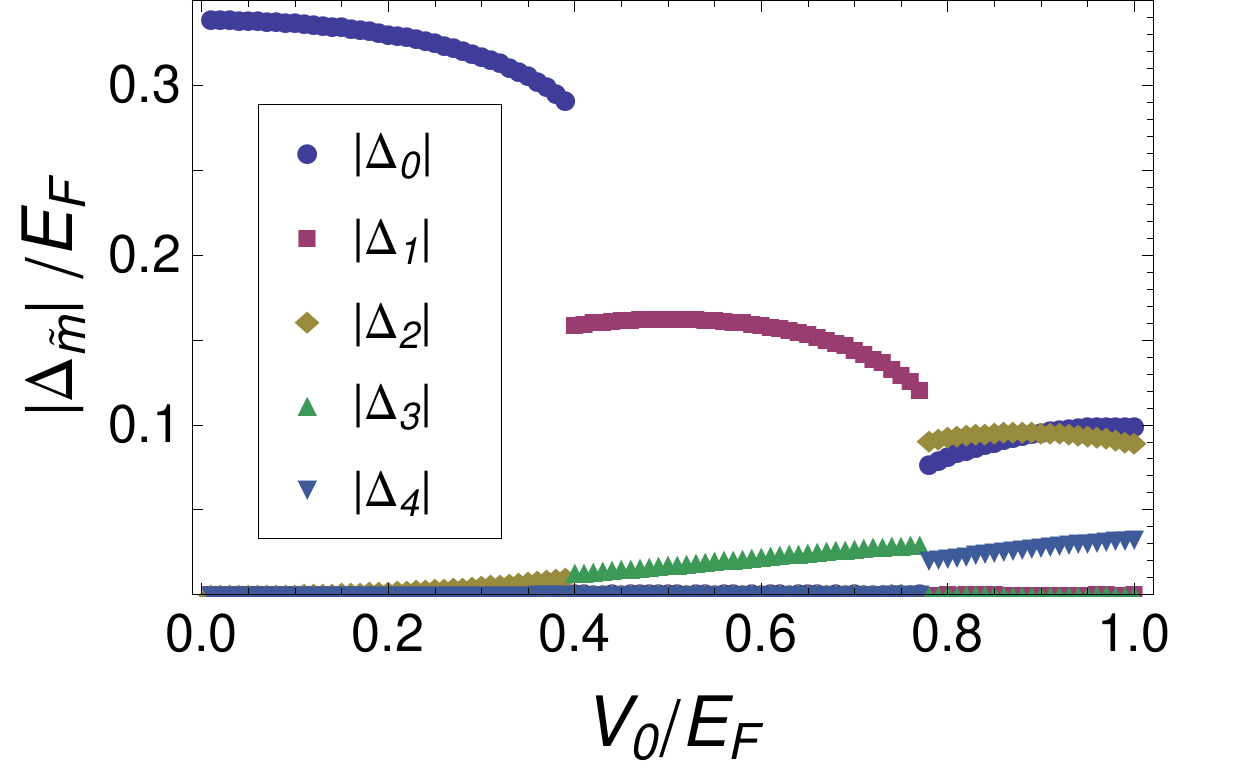}
\caption{Plot of the absolute value of the Fourier components of $\Delta(x)$ in Eq.~\eqref{eqnBdGSC} as function
of the lattice potential strength $V_0$ (parameters as in Fig.~~\ref{figGapDensities}).
Only even components are non-zero for $V_0/E_F\leq0.38$, whereas only odd ones are non-zero
for $0.39\leq V_0/E_F\leq 0.77$, and so forth.
This is the tell-tale sign of the transition from the 0- to the $\pi$-phase.}
\label{figAbsPsiTotalPolarPot}
\end{figure}

In Fig. \ref{figGapDensities} we show $(n_{\uparrow}(x)-n_{\downarrow}(x))/k_F$ and $\Delta (x)/E_F$ as functions of $2x/\lambda$
with $k_F\lambda=30$ for amplitudes $V_0/E_F=0.39$, 0.40, 0.77 and 0.78. Here we notice a sudden change in the gap profile from an even to an odd function (around $x=0$) at the definite value $V_0/E_F=0.39$, and again at $V_0/E_F=0.78$. We will give arguments as to why this occurs in the following sections.
The signature of the new phases are even more clear in Fig.~\ref{figAbsPsiTotalPolarPot} which shows the largest components of $|\Delta_{\tilde m}|$. Here we see a very clear jump between even and odd Fourier components as $V_0$ is increased. This transition constitutes the main result of our paper and below we propose a way to observe the $\pi$-phases which is clearly distinguishable from other oscillatory behaviors in the gap.

The spin-dependent lattice we use here is invariant under spatial reflection. 
In addition, for the balanced system we study, the symmetry of the lattice potential implies that the densities are even and interchanged every half wavelength ($\lambda /2$). We can use this observation to restrict the functional form of the gap.
In the absence of any currents, the gap obeys $\Delta(x+\lambda/2)=\pm\Delta(x)$. Combined with the full periodicity $\Delta(x+\lambda)=\Delta(x)$, we see that either only even or only odd Fourier components survive which facilitates its unmistakable detection.

Our spin-dependent lattice potential effectively acts as a spatially varying magnetic field. 
In contrast, in homogeneous systems the wavevector of a Fulde-Ferrell (FF) state increases monotonically with increasing chemical potential difference. Using the results presented in \cite{drummond2007}, we determine which chemical potential difference $V_0$ (at fixed density) is needed so that the order parameter of the FF state has the wavevector $2\pi/\lambda$ corresponding to the lattice used in our calculation. 
The result is presented as a dashed line in Fig.~\ref{figPhaseDiagram}. This line indicates where FF solutions exists, however only for $V_0\geq 0.27$ (the BCS-FF arrow in Fig.~\ref{figPhaseDiagram}) is it a global energy minimum.
The qualitative agreement clearly demonstrates the connection to FFLO physics.
The difference for our system is that the wavelength
of variations in the gap does not vary continuously but changes at discrete values of $V_0$ since the oscillations must be commensurate
with the lattice potential. FFLO states are similar to $\pi$-phases since modulation of the pairing field generates a lower energy solution. However, the present proposal differs from FFLO since we do not have a global spin imbalance and the system is intrinsically stable
to phase separation.

The transition between the different $\pi$-phases can be explained by an energy balance argument. They are driven by the competition between interaction (pairing and Hartree terms) and potential energy in the lattice.
First consider the situation where the gap and densities are almost constant and even functions, $\Delta^e(x)=\Delta_0$ and
$n_{\uparrow}(x)=n_{\downarrow}(x)=n_0/2$ ($V_0/E_F\leq 0.38$ in Fig~\ref{figGapDensities}), and
contrast this with the situation where the gap is an odd function. Let us for simplicity assume the same magnitude of the gap and
introduce a corresponding oscillation in the densities, thus $\Delta^o(x)=\Delta_0 \sin (2 \pi x /\lambda)$ and $n^{o}_{\uparrow/\downarrow}(x)=n_0/2\mp (\delta n/2) \cos (2 \pi x /\lambda)$. In the long-wavelength limit we can neglect the kinetic energy and  the energy densities of the even and odd state can be written
\begin{align}
v_{e/o}:= &\int dx \left[\sum_{\sigma}V_{\sigma}(x)n^{e/o}_{\sigma}(x)+\right.&\nonumber\\ &\left.g_{1D}n^{e/o}_{\downarrow}(x)n^{e/o}_{\uparrow}(x)+|\Delta^{e/o}(x)|^2/g_{1D}\right] /L,&
\end{align}
where $L$ is the system size. From our ansatz we get $v_{e} = g_{1D}n_{0}^{2}/4+\Delta_0^2/g_{1D}$
and $v_{o}= g_{1D}n_{0}^{2}/4+\Delta_0^2/(2 g_{1D})-V_0 \delta n/2-(g_{1D}/2)(\delta n/2)^2$. If we determine the
density variation in the odd state by requiring minimal energy, we find
$v_{o}= g_{1D}n_{0}^{2}/4+\Delta_0^2/(2 g_{1D})+V_0^2/(2g_{1D})$ from which it follows that the constant even solution is
lower in energy until $V_0=\Delta_0$.
Taking $\Delta_0/E_F \sim 0.3$ from Fig.~\ref{figGapDensities} gives $V_0/E_F\sim 0.3$.
Numerically we find $V_0/E_F\sim 0.39$, about 30\% higher.
At small $\lambda$ we expect large deviations from this estimate. This is caused by the neglected kinetic term that grows with decreasing $\lambda$ and pushes the jump to larger $V_0$. Non-commensurate solutions should be thermodynamically unstable by the same argument as they do not benefit from lowering of the potential energy. Initially we did search for non-commensurate solutions but as expected we found only commensurate ones.

\begin{figure}
\includegraphics[width=\columnwidth]{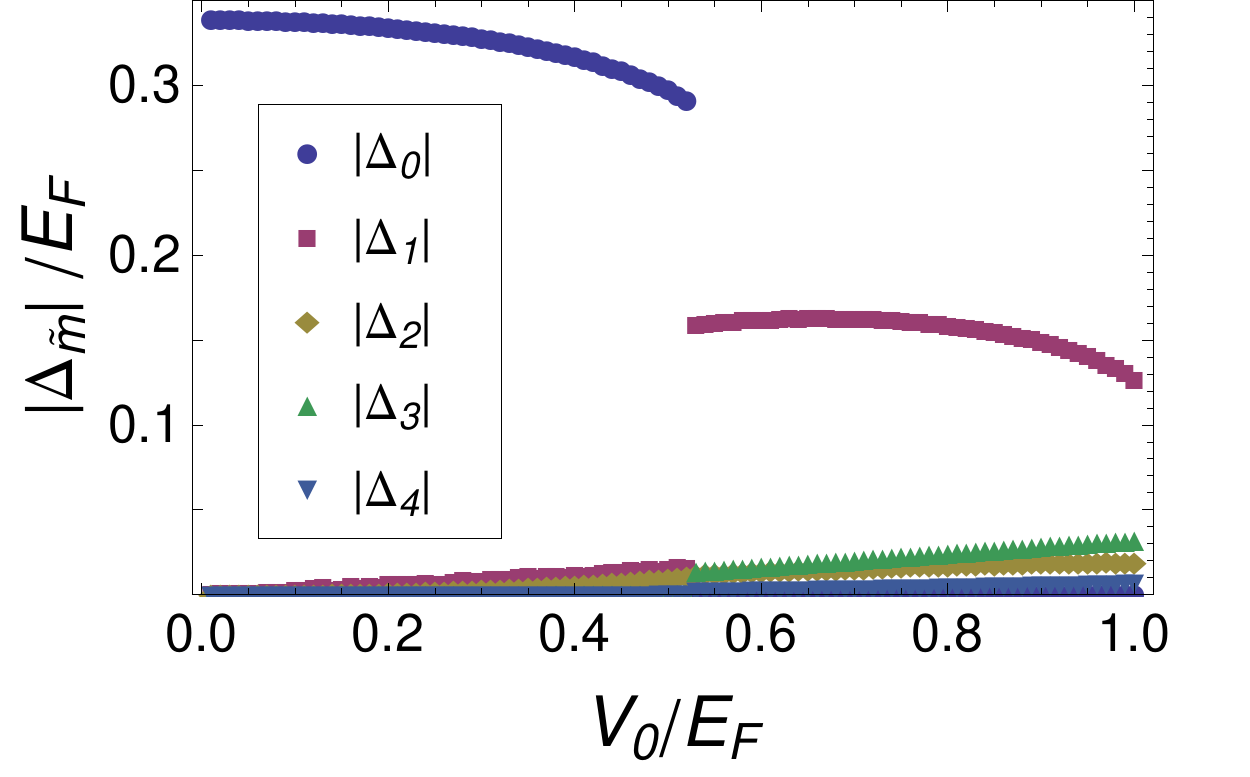}
\caption{Same as Fig. \ref{figAbsPsiTotalPolarPot} but with $V_{\uparrow}(x)=-3 V_{\downarrow}(x)/2=V_0 \cos (2 \pi x / \lambda)$.}
\label{figAbsPsiPartPolarPot1}
\end{figure}

The transitions we find are very sharp as illustrated in Figs.~\ref{figGapDensities} and \ref{figAbsPsiTotalPolarPot}. The $0$- and $\pi$-phase have
different parities and we therefore have a crossing of ground-states as we tune $V_0$.
We test the stability of our predictions by using a potential that has $V_\uparrow(x)=-3V_\downarrow(x)/2$. As Fig.~\ref{figAbsPsiPartPolarPot1} shows, a sharp transition occurs also in this case. Even though this potential breaks $\lambda/2$ symmetry, there is still conservation of parity, thus $\Delta_{-\tilde m}=\pm \Delta_{\tilde m}$, and a sharp transition still occurs. However, now the order parameter constains both even and odd Fourier components.

The results presented in Figs.~\ref{figAbsPsiTotalPolarPot} and \ref{figAbsPsiPartPolarPot1} have $k_F\lambda=30$.
For smaller $\lambda\sim \xi$, the lattice drives the system into the normal state before showing any noticeable oscillations of the gap. For the interaction strength $g_{1D}k_F/E_F=-2.04$, we estimate that $k_F\lambda\gtrsim 12$ is necessary to support observable $\pi$-phases. In the phase diagram in Fig.~\ref{figPhaseDiagram} the suppression of the gap at the transitions at small $\lambda$ is clearly seen. Fig.~\ref{figPhaseDiagram} also demonstrates that more zeros of the gap per unit cell could be accessible in experiments. The presented results are for the zero temperature case. 
For the parameter regime in Fig.~\ref{figPhaseDiagram}, we find that the phase diagram is qualitatively the same up to $T\sim0.1T_F$. We have not performed a detailed analysis of the transition to the normal state outside of this parameter regime or for even higher temperatures.
An interesting aspect of this system is that the spin-dependent potential leads to an appreciable triplet component in the pairing \cite{zap2010}.

In order to detect the $\pi$-phase, the rapid ramp technique can be used to transfer opposite spin pairs into molecules on the Bose-Einstein condensate (BEC) side of the Feshbach resonance as shown in \cite{greiner2003,regal2004,zwierlein2004}. We can calculate the number of molecules with center-of-mass momentum
$q:=2\pi \tilde m/\lambda$, with $\tilde m$ an integer, by projection of the state before the ramp into the bound $s$-wave molecular state $f_s(r)\sim \sqrt{2/a_s}\exp(-r/a_s)/r$, where $r$ is the relative distance and $a_s$ the scattering length on the BEC side \cite{diener2004,altman2005}.
Assuming $a_s,a_\bot\ll\xi$, with $a_\bot$ the transverse confinement length, we obtain $n_q\sim (8La_{\bot}^2/g_{1D}^{2}a_s)\kappa\left[a_\bot/a_s\right]^2 |\Delta_{\tilde m}|^2$ \cite{kappa}. 
Additional density terms give only a featureless continuous contribution to $n_q$. 
For $V_{\sigma}(x)=-V_{-\sigma}(x)$, $\pi$-phases are linked to either only odd or only even Fourier components of the order parameter. After the rapid ramp process, the BCS to $\pi$-phase transition is seen as the disappearance of a single spot and the apperance of two new spots, ruling out other types of oscillation in $\Delta$. In contrast, phase-separated densities located in the minima of the potentials would require many Fourier components. Observation of sharp peaks in the distribution is therefore a clear sign of $\pi$-phases.

For experimental realization we focus here on $^{40}$K \cite{moritz2005}. $^{6}$Li is another possible candidate, although we note that a spin-dependent lattice is harder to implement \cite{chin2006}.
We assume a 1D geometry of tubes that are optically trapped with a superposed magnetic field to control the interaction via the Feshbach resonance at $B_0=202.1$G. Using $N\sim 100$ per tube of length $L\sim 40\mu\text{m}$, we have $n\sim2.5$ $\mu$m$^{-1}$ and $k_F=\pi n/2\sim 3.93$ $\mu$m$^{-1}$. For simplicity we neglect the external confinement. With $a_\bot=60.3$nm and $a_s=132.3$nm from \cite{moritz2005}, and using $\Delta_{1}/E_F=0.15$ and $g_{1D}k_F/E_F=-2.04$, we find $n_{2\pi/\lambda}=1.58$. However, the number of tubes in \cite{moritz2005} was 4900 so we expect a signal of 7742 molecules, which should be detectable.

In the unpolarized system, the appearance of $\pi$-phases in the order parameter requires a spin-dependent lattice potential with a wavelength longer than the coherence length; $\lambda\gtrsim \xi$. To fulfil both requirements multiple lasers should be used \cite{jacksch1999, grynberg2001, liu2004}. To get spin-dependence there are several
proposals and we focus on the one of \cite{liu2004}.
The splitting is controlled by the difference in laser intensity and phase of left-circular and right-circular polarized light. Furthermore the transition is between the $^{2}$S$_{1/2}$ and $^{2}$P$_{1/2},^{2}$P$_{3/2}$ lines with optical wavelength $\lambda_{opt} \sim 770$ nm.
To change $\lambda$ in the lattice one changes the angle between laser and tubes. The magnetic field is aligned parallel to the lasers.
If $\theta$ is the angle between the tubes and these lasers, the lattice wave-length is $\lambda = \lambda_{opt} /2\cos (\theta)$. $k_F\lambda=30$ translates to 7.6 $\mu$m and $\theta \sim 87^o$. Since this is almost perpendicular to the
1D tube, the heating will also be reduced as most of the recoil is absorbed in the confining potential.

In an actual experiment it has been suggested that 1D tubes in an intermediate strength 2D optical lattice will give the best conditions for observing the FFLO state in 1D \cite{parish2007}, and we expect this to hold for our $\pi$-phases as well. Our proposal differs from other studies on FFLO states in cold atom since we use an unpolarized gas. The Fermi surfaces are therefore identical for the two spins and the non-trivial pairing properties of the system are entirely due to the spin-dependent lattice potential.

We thank R. Sensarma, D. Pekker, L. Fritz, D. Weld, M. A. Cazalilla and F. Sols for helpful discussions. The authors acknowledge support from Real Colegio Complutense en Harvard, MEC (Spain) grant FIS2007-65723, the German Research Foundation grant WU 609/1-1, the Villum Kann Rasmussen foundation, CUA, DARPA, MURI, and NSF grant DMR-0705472.

\end{document}